\documentclass[aps,prl,twocolumn,amsmath]{revtex4}
\usepackage{graphicx}
\usepackage{bm}
\usepackage{color}

\newcommand{\beq}{\begin{equation}}
\newcommand{\beqa}{\begin{eqnarray}}
\newcommand{\eeq}{\end{equation}}
\newcommand{\eeqa}{\end{eqnarray}}

\newcommand{\lmk}{\left(}
\newcommand{\rmk}{\right)}

\newcommand{\vc}{v_{\rm s}}


\begin{document}

\title{Search for an emission line of a gravitational wave background}

\author{Atsushi Nishizawa}
\email{anishi@caltech.edu}
\affiliation{Theoretical Astrophysics 350-17, California Institute of Technology, Pasadena, California 91125, USA}
\author{Naoki Seto}
\affiliation{Department of Physics, Kyoto University, Kyoto 606-8502, Japan}

\begin{abstract}
 In the light of the history of researches on electromagnetic wave spectrum, a sharp 
emission line of  gravitational-wave background (GWB) would be an interesting 
observational target. Here we study an efficient method to detect a line GWB 
by correlating data of multiple ground-based detectors. We find that the width of frequency bin for coarse graining is a critical parameter, and the commonly-used value 
0.25\,Hz is far from optimal,  decreasing the signal-to-noise ratio by up to a 
factor of seven. By reanalyzing the existing data with a smaller bin width, we might detect a 
precious line signal from the early universe.
\end{abstract}

\date{\today}

\maketitle

Newton refurbished our understanding of the colors of visible light by sophisticatedly utilizing prisms, in addition to his extraordinary accomplishments on gravitation and classical mechanics. His sequential works on optics were summarized in {\it Opticks} published initially in 1704. But we had to wait for another $\sim 100$ years for significant experimental developments on electromagnetic wave spectra. 
In 1800, Herschel uncovered infrared components of a light ray. Around the same time, Wollaston (more systematically Fraunhofer) discovered sharp absorption lines in the spectrum of Sun light. Moreover, the new element Helium was identified in the solar corona by its emission lines detected at the total solar eclipse in 1868. However, the physical mechanism itself behind the observed sharp
emission/absorption lines could be hardly clarified in 1800s. Indeed, 
this conundrum was one of the primary driving forces toward the 
construction of quantum mechanics in early 1900s.

On the other hand, the existence of a gravitational wave (GW) is predicted by 
Einstein's general theory of relativity that was completed in 1915. As the 
gravitational interaction is very weak, GWs have not been directly observed yet. However, with the advent of powerful second generation interferometric detectors, the first detection would be achieved shortly. 

The high penetrating power of GWs could become quite advantageous to probe the 
early universe. At present, our observational information 
before the big bang nucleosynthesis is severely limited \cite{Maggiore:1999vm}. Therefore, once primordial GWs are detected, they would provide us with invaluable clues to understand inflation and the subsequent reheating era \cite{Turner:1996ck,Boyle:2007zx,Watanabe:2006qe,Easther:2006gt,Nakayama:2008wy} and high energy physics \cite{Kamionkowski:1993fg,Grojean:2006bp,Damour:2004kw,Fenu:2009qf}. 
Also GW backgrounds (GWBs)  bring us opportunities to test gravity theories \cite{Seto:2006hf,Nishizawa:2009bf,Gumrukcuoglu:2012wt,Nishizawa:2014zra}. To maximally extract the scientific outputs from the accumulated data of the interferometers, we should deliberate on methods of data analysis and thoroughly search for stochastic GWBs, not only for given theoretical predictions but also in model independent manners.  

In the light of the aforementioned history of electromagnetic waves, it would be 
interesting to explore a sharp emission line of a cosmological GWB. In this letter, we study how a line GWB  looks like in detector signals, and suggest 
an efficient detection method by correlating data of multiple ground-based interferometers. We find that the width of frequency bin should be chosen carefully, and its preferred size is smaller than 0.1\,Hz. With the commonly used width 
$0.25$\,Hz \cite{Aasi:2014zwg}, the signal-to-noise ratio could be reduced by up to a factor of seven. Therefore, by reanalyzing the existing data, we might strike a 
precious line  signal that has been buried in detector noises.

{\it{Doppler broadening of GWs}} ---
Throughout this letter, GWB is assumed to be isotropic in the CMB rest 
frame. As a simple but suggestive example, we first examine a delta-function-like spectrum
concentrating solely at a frequency $f_{\rm r}$ (cosmologically generated at the 
same comoving wavelength). 
Our study can be 
easily extended to a GWB with multiple lines.
The effects of a line profile will be discussed later.

For a detector moving relative to the CMB frame with velocity $\vec{v}(t)$, the observed  Doppler-shifted frequency 
$f(t,\hat{\Omega})=f_{\rm{r}} \left[ 1-\vec{v}(t)\cdot \hat{\Omega} \right]$ is 
anisotropic and  
depends on the propagation direction of a GW 
$\hat{\Omega}$.
 In this letter,  we use the unit $c=1$ for the speed of light, and also drop 
 the terms of ${\rm O}(|\vec{v}(t)|^2)$ for the Doppler effect.

\begin{figure}[t]
\begin{center}
\includegraphics[width=7cm]{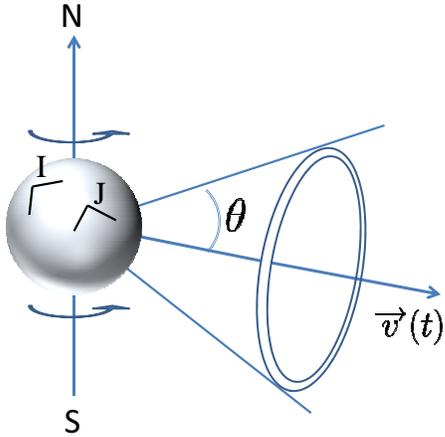}
\caption{Observation of a line GWB that is isotropic in the CMB rest frame. An 
observer (an  
interferometer on the Earth) moves at the relative velocity $\vec{v}(t)$, and sees an isotropic background due to the Doppler 
effect. The observed frequency is identical along the ring across the sky with 
$\theta=const$ ($\theta$: angle between $\vec{v}(t)$ and the propagation 
direction of GW $\hat{\Omega}$). The spin rotation of the Earth generates diurnal
modulation for the correlated signal of two interferometers $I$ and $J$.
}
\label{fig:aisotropic-GWB}
\end{center}
\end{figure}

The relative velocity $\vec{v}_{\rm{s}}$ of the Solar-System barycenter (SSB) to the CMB rest frame 
is toward  the Galactic coordinate $(\ell, b)=(263.99^{\circ} 
\pm 0.14^{\circ}, 48.26^{\circ} \pm 0.03^{\circ})$ with the magnitude 
$\vc\equiv|\vec{v}_{\rm{s}}|=(1.230 \pm 0.003)\times 10^{-3} $ (corresponding 
to $369.1{\rm{km}}\,{\rm{s}}^{-1}$) \cite{Aghanim:2013suk}.
Below, for simplicity, we put $\vec{v}(t)=\vec{v}_{\rm s}$, neglecting  corrections 
(at most $\lesssim$10\%) due to the velocity of the detector relative to the SSB.
We also neglect   minor  relativistic effects such as the Sachs-Wolfe fluctuations. These would be  reasonable 
approximation for our demonstration, and we can easily employ more 
sophisticated models in 
actual data analysis.  
Then we have
\begin{equation}
f(t,\hat{\Omega}) = f_{\rm{r}} \left[ 1-v_{\rm s}\cos \theta \right]
\label{eq12}
\end{equation}
with the angle $\theta$ between $\hat{\Omega}$ and $\vec{v}_{s}$.
For a given  observational frequency,  
the 
GW signals come from a ring on the sky as shown in 
Fig.~\ref{fig:aisotropic-GWB}.
The total width of the Doppler broadening 
 is given by
\beq
\delta f_{\rm{D}} =2f_{\rm r}v_{\rm s}\sim 0.2 \lmk 
\frac{f_{\rm{r}}}{100\,{\rm{Hz}}}\rmk\,{\rm{Hz}}. \label{dop}
\eeq

The optimal band of  ground-based interferometers is around $10$-$1000$\,Hz. 
Meanwhile, by taking Fourier 
transformation for an appropriate time-segment (duration $T$) of data, we can realize a fine frequency resolution $\sim 10^{-3} 
(T/1000\,{\rm sec})^{-1}$\,Hz, unlike an electromagnetic-wave spectrum. But, as we discuss later,  the  spin rotation of the Earth plays an 
important role and this limits the duration $T$ for the coherent Fourier transformation.

{\it{GW signal}} --- 
Next, we briefly summarize the correlation analysis for an anisotropic GWB (not 
specific to a line GWB) \cite{Allen:1996gp}. 
In the frequency space, the GW signal of a detector at the position $\vec{X}_I$ 
can be generally expanded as 
\begin{equation}
h(f,\vec{X}_I) = \sum_A \int _{S^2} d\hat{\Omega}\, \tilde{h}_A (f, \hat{\Omega}) e^{-2\pi i f \hat{\Omega} \cdot 
\vec{X}_I}\, F^A_I (\hat{\Omega})\:.
\label{eq2}
\end{equation}
Here, the label $A$ denotes the polarization mode ($A=+, \times$) of a GW, and $\tilde{h}_A (f, \hat{\Omega})$ is the Fourier mode of the GW  in each polarization 
state. The factor  $F^A_I$ is the antenna pattern function of the $I$-th GW 
detector, and has a quadrupolar angular dependence.
The power spectrum $S_h(f)$ for $\tilde{h}_A (f, \hat{\Omega})$ is given by the 
energy density  of GWB $\Omega_{\rm{gw}}(f; \hat{\Omega})$ (per logarithmic 
frequency interval per steradian) normalized by the critical density of the 
universe as
\begin{equation}
\Omega_{\rm{gw}}(f; \hat{\Omega}) = \frac{4\pi^2}{3H_0^2} f^3 S_h(f; \hat{\Omega}) \;,
\end{equation}
where $H_0$ is the Hubble parameter and we use $H_0= 
70\,{\rm{km}}\,{\rm{s}}^{-1}\,{\rm{Mpc}}^{-1}$.

The correlation analysis  is an efficient method to  distinguish a GWB signal 
from detector noises (see {\it e.g.}  \cite{Christensen:1992wi,Flanagan:1993ix,Allen:1997ad}). 
In this method, we prepare data streams of two widely-separated 
detectors (with independent noises), and  take cross correlations of their   Fourier modes.  For statistical amplification and coarse graining, we divide the Fourier space into multiple bins.  Below, a bin is specified by the label $k$, and 
its  central frequency and width are denoted  by $f_k$ and $\delta f_b$, 
respectively.  We intrinsically have $\delta f_b>T^{-1}$ due to the frequency resolution.  
Within each bin  $k$,  we take a summation of the correlated products.  Its expectation value  is written as \cite{Allen:1997ad}
\beq
\mu_k = \frac{3H_0^2}{8 \pi^2} \frac{T \delta f_b}{f_k^3} Z (f_k). \label{eq6} 
\eeq
Here we defined the integral
\beq
Z (f_k) \equiv \int \frac{d\hat{\Omega}}{4\pi} \Omega_{\rm{gw}} (f_k; \hat{\Omega}) \gamma(f_k,\hat{\Omega}) \;, \label{eq14}
\eeq
and the direction-dependent overlap-reduction function
\begin{equation}
\gamma(f,\hat{\Omega}) \equiv \sum_{A} F^A_I (\hat{\Omega}) F^{A}_J (\hat{\Omega}) \exp \left[ 2 \pi i f \hat{\Omega} \cdot \vec{d}_{IJ} \right] \;
\label{eq13}
\end{equation}
with $\vec{d}_{IJ}(t) \equiv \vec{X}_I(t) - 
\vec{X}_J(t)$ for  two detectors $I$ and $J$. At high frequency regime, the phase factor $\exp \left[ 2 \pi i f \hat{\Omega} \cdot 
\vec{d}_{IJ} \right]$ depends strongly on the propagation direction 
$\hat{\Omega}$. In Eq.(\ref{eq6}),
we  assumed to take a small  bin width $\delta f_b$ so that the frequency 
dependence of relevant functions can be neglected  within each bin. 
But this is not always the case, as we later discuss in detail.

Now we apply the above expressions specifically to the line GWB at a 
frequency $f_{\rm r}$ (defined in the CMB frame).
From Eq.(\ref{eq12}), we can write the observed spectrum by
\beq
 \Omega_{\rm{gw}}(f;\theta) = {\tilde{e}_{\rm{gw}}}{f_{\rm r}}\ \delta 
 [f-f_{\rm{r}} (1-v_{\rm s} \cos \theta)],
\eeq
where $\tilde{e}_{\rm{gw}}$ is the total energy of the line normalized by the 
critical density.

For a line GWB, we  define the overlap-reduction function 
$\Gamma (f,u) $ that is obtained by
integrating $ \gamma(f,\hat{\Omega})$ across the ring directions at $u\equiv \cos\theta={\rm{constant}}$ as
\begin{equation}
\Gamma (f,u) \equiv \int_0^{2\pi} \frac{d \phi}{2\pi} \gamma(f,\hat{\Omega}) \;.  \label{eqg}
\end{equation} 
This definition is different from the following standard one $\gamma_{\rm st}(f)$ by an additional $u$-integral and 
the overall normalization; 
\beq
\gamma_{\rm{st}} (f) \equiv \frac{5}{8\pi} \int_{-1}^1du\int_0^{2\pi} {d \phi} \gamma (f,\hat{\Omega}) \;. \label{st} 
\eeq
From Eqs.(\ref{eq14})-(\ref{eqg})
we have
\beq
Z (f_k) = \frac{\tilde{e}_{\rm{gw}}f_{\rm r}}{2} \int_{-1}^1 du\, \delta [f_k-f_r (1-\vc u)]\, \Gamma (f_k,u) \nonumber \;.
\eeq
Replacing the frequencies of slowly-varying functions with $f_{\rm{r}}$ and 
putting $\Gamma(f_k)\equiv \Gamma(f_{\rm r},u(f_k))$, 
we obtain
\begin{equation}
\mu_k = \frac{3H_0^2}{16 \pi^2} \frac{T \tilde{e}_{\rm{gw}}}{f_r^3 \vc}~ \delta f_b ~\Gamma (f_k) \;. \label{eq11}
\end{equation}
Note that  the frequency $f$ and the sky angle $\theta$ (equivalently $u$) are related by Eq.~(\ref{eq12}). 

\begin{figure}[t]
\begin{center}
\includegraphics[width=8cm]{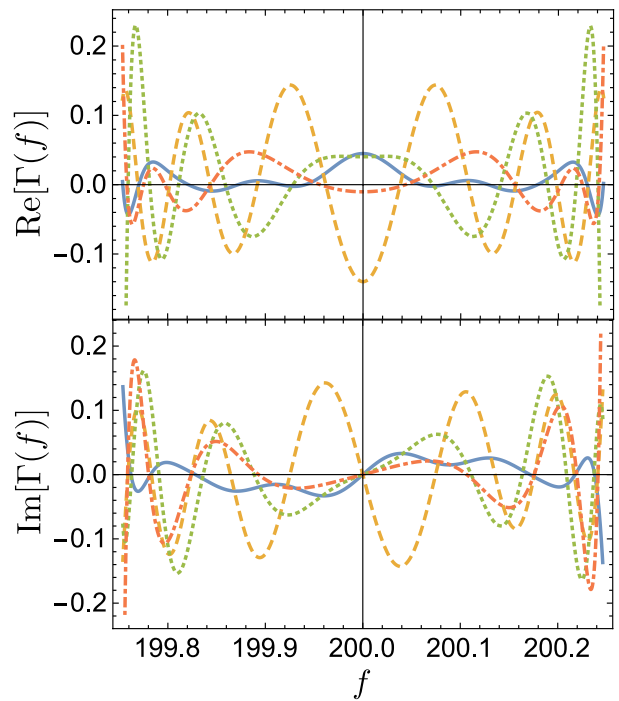}
\caption{Overlap-reduction function $\Gamma(f)\equiv \Gamma(f_{\rm r},u(f))$ for two LIGO interferometers with $f_{\rm{r}}=200\,{\rm{Hz}}$. 
The observed frequency $f$ and the propagation direction $u\equiv \cos\theta$ 
are related by Eq.~(\ref{eq12}).  
The upper panel is the real part and the lower panel is the imaginary part. The 
overlap-reduction function is modulated by the spin rotation of the Earth. Its shape is plotted at  four different epochs; the fiducial time (blue, 
solid), 3h later (orange, dashed), 6h later (green, dotted), and 9h later (red, 
dot-dashed). The characteristic frequency interval $\delta f_c\sim 0.1$\,Hz of the 
oscillations
is given by Eq.(\ref{fcc}).}
\label{fig:Gamma}
\end{center}
\end{figure}

In Fig.~\ref{fig:Gamma},  we show the directional overlap-reduction functions 
$\Gamma(f_k)$   observed by the two  advanced-LIGO (aLIGO)  
interferometers for a line GWB at $f_{\rm r}=200$\,Hz.  The frequency 
dependence changes diurnally due to the spin rotation of the Earth.

The left and right edges of the curves are for GWs coming from 
the cold and hot spots of the CMB dipole, and  have large amplitudes. This is  
because the GW signals are hardly canceled  by the $\phi$-integral 
in Eq.(\ref{eqg}).  In Fig.2, we should also notice that the real (imaginary) 
part is symmetric  
(anti-symmetric) around the intrinsic frequency $f_{\rm{r}}$. In relation to 
this structure, 
for an isotropic GWB ({\it e.g.} without a line component), the imaginary parts 
of the correlated signals vanish due 
to a cancellation. In contrast, for a line GWB, the
 overlap-reduction function  $\Gamma(f)$ can have finite imaginary parts because of the anisotropy induced by 
 the Doppler effect. 

In Fig.2, the wavy structure of the function $\Gamma(f)$ is mainly determined by the phase factor in Eq.~(\ref{eq13}). An oscillation occurs when the phase changes by $\delta \alpha \approx 2\pi$, where $\alpha \equiv 2\pi f_{\rm{r}} d_{IJ} \cos \beta$ and $\cos \beta \equiv \hat{\Omega}\cdot \hat{d}_{IJ}$. 
We also have $\delta (\cos\beta)\sim \delta(\cos \theta)\sim 1/(f_{\rm r}d_{IJ})$. 
Then the characteristic frequency interval $\delta f_c$ for the oscillation is 
evaluated as
\beq
\delta f_c\sim f_{\rm r}\vc\delta(\cos\beta)\sim \frac{\vc}{d_{IJ}}\simeq 
0.1\lmk \frac{d_{IJ}}{3000\,{\rm{km}}} \rmk^{-1}{\rm Hz}.\label{fcc}
\eeq
Comparing this result with the  Doppler width (\ref{dop}), 
the number of oscillations becomes larger 
for a higher frequency $f_{\rm r}$. 
  For another detector pair more separated than aLIGOs ($d_{IJ}\sim$3000\,{\rm{km}}), the characteristic 
interval $\delta f_c$ becomes smaller.

{\it{Signal-to-noise ratio}} ---
The correlated signal (\ref{eq11}) would be suitable for a matched filtering 
analysis with the two fitting parameters $f_{\rm r}$ and $\tilde{e}_{\rm gw}$.
To calculate the signal-to-noise ratio (SNR) of the line detection, we evaluate 
the  
variance of the correlated signal $\mu_k$. Here, for each Fourier mode,  the noises of the 
two detectors are assumed to be independent and to have magnitudes  much 
larger than those of the GW signals. Then, the variance of the correlation 
signal $\mu_k$ is mainly contributed by detector noises and is given by \cite{Allen:1997ad}
\begin{equation}
\sigma_k^2 \approx \frac{T}{8} \delta f_b \,P_I(f_k) P_J(f_k) \;, \label{eq7}
\end{equation}
with   the one-sided noise spectra $P_I(f)$ and $P_J(f)$. 

The correlation signal in Eq.~(\ref{eq6}) is a complex number, as mentioned 
before. From Eqs.~(\ref{eq6}), (\ref{eq11}), and (\ref{eq7}), the total squared 
SNR is evaluated as
\beq
\rho^2  
=\sum_k\frac{|\mu_k|^2}{\sigma_k^2}
\approx \frac{9H_0^4}{32 \pi^4} \frac{T \tilde{e}^2_{\rm{gw}}\delta f_b }{f_r^6 v_s^2 P_I(f_r) P_J(f_r)} \sum_k \left| \Gamma (f_k) \right|^2 \;.
 \label{eq21}
\eeq
For a total observational duration $T_{\rm obs}$ much longer than a day, we need to divide the data 
into time segments for taking Fourier transformation.   But, in Eq.(\ref{eq21}), 
 we can effectively
regard $T$  as the duration $T_{\rm obs}$, after replacing $ \left| \Gamma (f_k) \right|^2 $  by its time average. 
Importantly,  the total SNR $\rho$ does not 
depend on the details of the time segmentation, once we can take a sufficiently small bin $\delta 
f_b$ to resolve the function $\Gamma(f_k)$.

{\it{Signal cancellations within bins}} ---
So far, we have assumed that the bin width $\delta f_b$ is much smaller than the 
characteristic frequency interval $\delta f_c$. If this is not the case, the 
coarse-grained signal $|\mu_k|$ decreases due to the averaging of the wavy structure around zero point, as 
understood from Fig.2.  Then, for Eq.(\ref{eq21}),  we should use the following integral 
\begin{equation}
\mu_k = \frac{3H_0^2}{16 \pi^2} \frac{T \tilde{e}_{\rm{gw}}}{f_r^3 \vc} \int_{\delta f_b} df \, \Gamma (f) \;,
\label{eq10} 
\end{equation}
in stead of the original one (\ref{eq11}) that is valid only when the variation 
of the function 
$\Gamma(f)$  
 is negligible within each bin. 
Therefore, to realize a high SNR,  we should set the bin width sufficiently 
smaller than the characteristic interval
$\delta f_c\sim 0.1$\,Hz.  Nevertheless, the commonly used value is $\delta f_b=0.25$\,Hz, e.g.~in \cite{Aasi:2014zwg}, and the sensitivity to a line GWB might be decreased significantly. 
Below, we take a close look at this degradation.

\begin{figure}[t]
\begin{center}
\includegraphics[width=8.5cm]{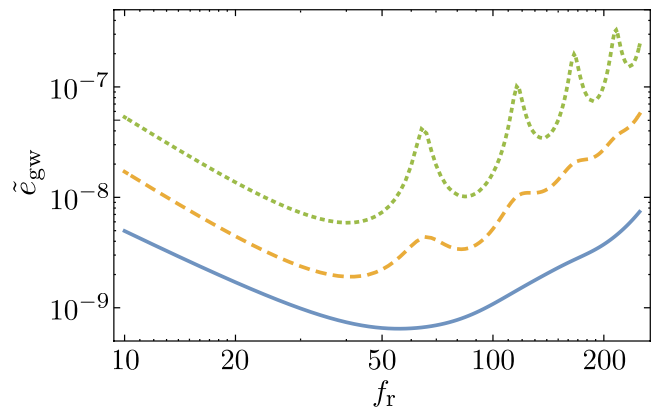}
\caption{Sensitivity to the normalized energy density $\tilde{e}_{\rm{gw}}$ of a 
line GWB. We assumed a  $1\,{\rm{yr}}$ observation with two aLIGO detectors and the 
detection threshold of $\rho= 10$. The lower curve (blue, solid) is for the bin 
width $\delta_b=0$\,Hz. The middle (orange, dashed) and upper (green, dotted) are for the bin widths of $0.25\,{\rm{Hz}}$ and $2.5\,{\rm{Hz}}$, respectively.}
\label{fig:sensitivity}
\end{center}
\end{figure}

{\it{Sensitivity to a line GWB}} ---
In Fig.~\ref{fig:sensitivity}, the sensitivity to the normalized energy density  
$\tilde{e}_{\rm{gw}}$  is shown as a function of the line frequency $f_{\rm r}$.
Here, we assumed a  $1\,{\rm{yr}}$ observation with two aLIGO detectors at the 
detection threshold of $\rho= 10$. 
For their noise  spectra $P_{I,J}(f)$, we use the fitting formulas given in \cite{Sathyaprakash:2009xs}. 

In our study, the key parameter is the width of the frequency bin $\delta f_b$.
 For our demonstration here, we selected the following three widths;  $0$\,Hz, 
 $0.25\,{\rm{Hz}}$, and $2.5\,{\rm{Hz}}$. The first one means
a sufficiently small width (still satisfying $\delta f_b>T^{-1}$).
We use Eq.~(\ref{eq21}) for this one and Eq.~(\ref{eq10}) for the other two.
Because, for a finite width, the sensitivity 
 depends on the actual positions of the bins, we plotted the sensitivity 
 averaged  over  their possible positions (by  shifting boundaries of bins).

   In Fig.3, the sensitivity decreases 
 significantly for larger bin widths.  This can be expected from the previous 
 arguments on the cancellation of the wavy structure. 
The degradation of the sensitivity, say, at $f_{\rm{r}}=70\,{\rm{Hz}}$, is by 
factors of 6.8 for $\delta f_b=0.25\,{\rm{Hz}}$  and 22 for $2.5\,{\rm{Hz}}$. Note that these ratios are independent of the noise spectra. 
We also examined bin widths smaller than $\delta f_c=$0.1\,Hz.
For a line GWB with $10{\rm Hz}<f_{\rm r}<600{\rm Hz}$, the degradations 
(compared with $\delta f_b=0$) 
are within 1.1 for $\delta f_b=0.01$\,Hz, 1.2 for $0.02$\,Hz,  1.5 for $0.04$\,Hz, and 3.5 
 for $0.09$\,Hz. For example, to keep the loss of the sensitivity within 20\%, we 
 should take a bin width $\delta f_b$ smaller than 0.02\,Hz, in contrast to the 
 standard  choice 0.25\,Hz.

In the opposite limit with  $\delta f_b\gg \delta f_{\rm D}$, the $f$-integral in
 Eq.(\ref{eq10}) is equivalent to posing an additional $u$-integral to Eq.(\ref{eqg}). The resultant  is proportional to the standard overlap-reduction function $\gamma_{\rm st}(f)$ defined in Eq.(\ref{st}).  Indeed, in Fig.3, the positions of the peaks are around the zero points of the standard one $\gamma_{\rm st}(f)$ \cite{Allen:1997ad}. 

{\it{Discussions}} --- 
In the analysis above, we assumed a delta-function-like profile for a line 
GWB.  Here we briefly comment on the effects of the intrinsic line width  
$\Delta f_{\rm{gw}}$ (now only considering $\delta f_b\ll \delta f_c$). 
This  width $\Delta f_{\rm gw}$ should be compared with the Doppler 
broadening $\delta f_{\rm{D}}$ and the characteristic frequency interval $\delta f_c$.
For a sharp line with $\Delta f_{\rm gw}\ll \delta f_c$, we can apply our previous
arguments without change.

In the limit of a broad spectrum with $  \Delta 
f_{\rm{gw}}\gg  \delta f_{\rm{D}}$, a GWB at a certain observed frequency is contributed from 
the whole sky directions by the Doppler effect. There is no correspondence between an 
observed frequency $f$ and the angle $\theta$. In this case, by replacing $\Omega_{\rm{gw}}(f_k;\hat{\Omega})$ with $\Omega_{\rm{gw}}(f_k)$
and using  Eqs.~(\ref{eq14}), (\ref{st}) and (\ref{eq21}),
we have 
\beq
\rho^2 = \frac{9H_0^4}{50 \pi^4} T \delta f_b \sum_k  \frac{\Omega_{\rm{gw}}^2(f_k) \gamma_{\rm{st}}^2(f_k)}{f_k^6 P_I(f_k) P_J(f_k)}\;. \eeq
This is the standard SNR formula for a broad-band isotropic GWB 
\cite{Allen:1997ad}. Note that, for a fixed total energy density $\int d{\ln}f\,
\Omega_{\rm gw}(f)={\rm{constant}}$ ({\it e.g.} for the bound from the big bang 
nucleosynthesis \cite{Maggiore:1999vm} or CMB \cite{Smith:2006nka}), we can readily derive an approximate scaling relation $\rho\propto (\Delta 
f_{\rm gw})^{-1/2}$ for the SNR.

 In intermediate case with $\delta f_c < \Delta f_{\rm{gw}} < 
 \delta f_{\rm{D}}$, a GW signal at a certain observed frequency is contributed 
by GWs from a band region in the sky. Thus the GW 
 signals are partially canceled, and the SNR is smaller than that for $\Delta 
 f_{\rm{gw}}=0$ (for a fixed energy density).

When searching for a line GWB with real detectors, one might worry 
about confusion with various noise lines  produced by instruments. 
However, we can clearly distinguish them by using their time 
dependences. In contrast to noise lines, a GW line is modulated due to the rotation and revolution of 
the Earth, showing the  characteristic profiles as illustrated in Fig.2. Therefore, by 
appropriate Matched filtering analyses, one could easily distinguish the   
true GW signals from artificial noises.

Although we have only considered aLIGO pair, our study can be also 
applied to other planned detectors, such as LISA (sensitive around $10^{-4}$-$10^{-1}$\,Hz) and Einstein Telescope (ET) ($1$-$1000\,{\rm{Hz}}$).
With the original LISA of the three-arms configuration \cite{LISA-report-1998}, we can construct two noise-independent data streams 
\cite{Prince:2002hp}. But, due to the symmetry of the system, 
their correlation identically vanishes for an isotropic GWB (as usually assumed).  In contrast, a line GWB appears as a strongly anisotropic signal and could be 
detected, in principle, by the correlation. The situations are similar for ET with the proposed three-arm
 configuration \cite{Punturo:2010zza}.
However, eLISA is currently planned to have just  two arms and can generate only single data stream  
\cite{Seoane:2013qna}.  
 In this case, our correlation method no longer works. For a space mission like 
 LISA, its third arm is 
 essential to explore a line GWB.   

{\it{Conclusions}} --- 
We have studied the method for efficiently searching for a line GWB. The sensitivity might be  degraded significantly, unless the bin 
width $\delta f_b$  is much smaller than 0.1\,Hz (in the case of two 
LIGO pair).   Nevertheless, for correlation analysis of GWBs,  the commonly used 
width is $0.25\,{\rm{Hz}}$.
By reanalyzing the existing data with a smaller width, we might actually uncover an important signal 
from the  early universe.  

\begin{acknowledgments}
A. N. is supported by JSPS Postdoctoral Fellowships for Research Abroad.
N.S.  is supported by JSPS (24540269) and
MEXT (24103006).
\end{acknowledgments}


\bibliography{/Volumes/USB-MEMORY/my-research/bibliography}

\end{document}